\documentclass{pasj01}
\usepackage{graphicx}	% Including figure files
\usepackage{natbib}
\usepackage[switch,mathlines]{lineno}

\def\ps{$\rm km \,s^{-1}\,kpc^{-1}$}
\def\arcdeg{$^\circ$}

\Received{2025/Feb/20}%{yyyy/mm/dd}
\Accepted{2025/May/22}%{yyyy/mm/dd}
\newcommand{\nar}{New Astron. Rev.}
\begin{document} 

%\linenumbers

\title{ 
Influence of Bar Formation on Star Formation Segregation and Stellar Migration: Implications for Variations in the Age Distribution of Milky Way Disk Stars
}

%%% begin:list of authors
\author{Junichi \textsc{Baba}\altaffilmark{1,2}}
\email{babajn2000@gmail.com; junichi.baba@sci.kagoshima-u.ac.jp}

\altaffiltext{1}{Amanogawa Galaxy Astronomy Research Center, Graduate School of Science and Engineering, Kagoshima University, 1-21-35 Korimoto, Kagoshima 890-0065, Japan.}
\altaffiltext{2}{Division of Science, National Astronomical Observatory of Japan, Mitaka, Tokyo 181-8588, Japan.}

%% `\KeyWords{}' always has to be placed before ``\maketitle'' 
%%  List of Key Words:  https://academic.oup.com/pasj/pages/Pasj_Keywords 

\maketitle

\begin{abstract}
We present a chemo-dynamical $N$-body/hydrodynamic simulation of an isolated Milky Way-like galaxy to investigate how bar formation influences star formation rates, stellar migration, and the resulting age and metallicity distributions of disk stars. Focusing on the transient epoch of bar formation, a phase that triggers gas inflows, enhances local star formation, and drives significant orbital migration, we find that the star formation rate in the inner disk exhibits a pronounced peak during this period. This behavior arises from the combined effect of vigorous star formation driven by strong spiral arms prior to bar formation and the subsequent suppression of star formation once the bar is established. In contrast, star formation in the outer disk persists after bar formation at modest levels, and enhanced outward migration of stars originally formed in the inner regions gives rise to a pronounced peak in the outer disk’s stellar age distribution corresponding to the bar formation epoch. Moreover, stars formed during this epoch tend to exhibit higher gas-phase metallicities, reflecting their origin in more metal-rich inner regions. Although our model does not capture every detail of the Milky Way's complex evolution, our results highlight the dominant role of bar driven migration in segregating star formation activity and in shaping the long-term chemical and age structure of the Galactic disk. Recent observational studies suggest that the Milky Way's bar is approximately 8 Gyr old; therefore, our findings imply that the age distribution of stars in the solar circle and outer disk should show a corresponding peak around that age. 
\end{abstract}

%\pagewiselinenumbers

%%%%%
\section{Introduction}

Galactic bars are a common feature in disk galaxies and play a critical role in their overall evolution \citep[e.g.][]{Sellwood2014reviw}. Early $N$-body/hydrodynamic simulations demonstrated that bar formation strongly enhances radial mixing and redistributes angular momentum, thereby affecting both the dynamical and chemical evolution of the disk \citep[][]{FriedliBenz1995,Friedli+1994}. By driving gas inflows into central regions \citep[][]{Athanassoula1992a,Li+2015,Sormani+2018a,Tress+2020} and modulating star formation across the disk, bars reshape metallicity gradients and influence the overall enrichment history of galaxies \citep[][]{Kubryk+2013,Grand+2015}. Numerous numerical and theoretical studies have shown that bars induce significant structural and kinematic changes, altering star formation patterns in both the central and inner parts of the disk \citep[e.g.,][]{Spinoso+2017,Sormani+2020,Fragkoudi+2020,BabaKawata2020a,Baba+2022} and driving strong orbital migration of stars and gas \citep[][]{DiMatteo+2013,Khoperskov+2020}.

In the Milky Way, infrared observations revealed that the bulge exhibits a boxy or peanut shape, suggesting the presence of a bar \citep[][]{BlitzSpergel1991a,Nakada+1991}. Follow-up photometric and spectroscopic studies have refined our view of the Galactic bar/bulge, showing a boxy/peanut-shaped bulge that transitions into a thinner bar approximately 5 kpc in length \citep[][]{WeggGerhard2013,Wegg+2015,Anders+2019}. These observations further indicate that the bar’s major axis is oriented at roughly 25$^\circ$ relative to the Sun-Galactic center line, and kinematic analyses estimate a bar pattern speed of about 35-40 \ps{} \citep[for reviews,][]{Bland-HawthornGerhard2016,HuntVasiliev2025}. Moreover, studies of the stellar age distributions in the nuclear stellar disk and the boxy/peanut bulge suggest that the Galactic bar formed roughly 8 Gyr ago \citep[][]{Bovy+2019,Nogueras-Lara+2020,Schodel+2023,Sanders+2024}. This implies that star formation activity and the efficiency of stellar migration in the Milky Way disk may have undergone rapid changes in the past as a result of bar formation. Indeed, recent observational studies discusses the bar’s role in redistributing angular momentum, altering radial abundance gradients, and shaping the age and metallicity distributions across the Milky Way disk \citep[][]{Nepal+2024,Haywood+2024}.

Recent large-scale surveys of the Milky Way have considerably advanced our understanding of its metal composition and stellar age distribution. Observational studies consistently report variability in the stellar age distribution \citep[][]{Snaith+2015,Mor+2019,Ruiz-Lara+2020,Sahlholdt+2022,del-Alcazar-Julia+2025}, yet there is no consensus on the precise locations of the observed peaks or the underlying causes of these variations. For example, \citet{Ruiz-Lara+2020} found that the age distribution of stars within about 2 kpc of the solar neighborhood shows sharp peaks at roughly 1 Gyr, 1.7 Gyr, and 7 Gyr, suggesting that pericentric passages of the Sagittarius dwarf galaxy have perturbed the Milky Way and triggered bursts of star formation. In contrast, \citet{Snaith+2015} argued that there is a pronounced dip around 8 Gyr, which they attribute to a rapid decline in star formation \citep[``bar quenching'';][]{Haywood+2016a}. In addition, \citet{Haywood+2024} analyzed stellar age and chemical composition data over galactocentric distances from 2 to 20 kpc and discovered a break in the radial [Fe/H] profile of stars aged 7 to 8 Gyr near 10 kpc. They interpret this feature as an effect of the bar's outer Lindblad resonance, suggesting a bar age of roughly 7 to 8 Gyr.

Nevertheless, it remains unclear whether stellar age distributions derived from regionally limited samples accurately trace the true star formation history of the Milky Way. The Milky Way hosts spiral arms and a bar, non-axisymmetric structures that drive stellar orbital migration, so that stars may have been born in locations different from where they are presently observed \citep[e.g.,][]{SellwoodBinney2002,Roskar+2008a,MinchevFamaey2010,Grand+2012b,Minchev+2014b,Halle+2015,Lu+2024}. Therefore, the stellar age distribution does not necessarily reflect the underlying star formation history.

Against this background, our study aims to clarify how bar formation influences the star formation history of the Milky Way disk by triggering changes in both star formation activity and stellar migration. In this paper, we present a chemo-dynamical $N$-body/smoothed particle hydrodynamic (SPH) simulation of an isolated Milky Way-like galaxy. We focus on the transient epoch of bar formation, a phase that triggers gas inflows, boosts inner disk star formation via strong spiral arms, and then suppresses star formation once the bar is established. Our simulation shows that the interplay between these processes produces a pronounced peak in the stellar age distribution of the outer disk. In particular, the enhanced outward migration of stars formed in the inner disk prior to bar formation results in an outer disk population that is more metal rich than the gas at their current locations, reflecting their origin in the inner regions.

Although our model does not capture every detail of the Milky Way's complex evolution, our results offer a conceptual framework for understanding how bar-induced secular evolution can drive major shifts in the star formation history and chemical structure of the Galactic disk. Our study is intended not to reproduce all observed variations in the star formation history, but to illuminate the role of bar formation in driving stellar migration and segregating star formation activity over cosmic time.

%%%%%
\section{Models and Methods}
\label{sec:ModelMethod}

To investigate how bar formation affects the chemical evolution of a galactic disk, we performed a chemo-dynamical simulation of an isolated galaxy. We adopted structural parameters from the Milky Way model of \citet{McMillan2017} and used {\tt AGAMA} \citep{Vasiliev2019} to generate the initial axisymmetric configuration. This setup includes live thin and thick stellar disks, a gaseous disk, a classical bulge, and a dark matter (DM) halo. In this model, the gas disk has a central hole at $R \lesssim 2\,\mathrm{kpc}$. By focusing on an isolated galaxy, we can specifically examine the role of bar formation without the added complexity of cosmological accretion or environmental effects.

We used the $N$-body/smoothed particle hydrodynamic simulation code {\tt ASURA-3} \citep[][]{Saitoh2017}. Hydrodynamics are computed using the density-independent SPH (DISPH) method \citep[][]{SaitohMakino2013}, which allows us to properly treat contact discontinuities and fluid instabilities. We computed self-gravity for all particles via a Tree with GRAPE method using the Phantom-GRAPE software emulator \citep{Tanikawa+2013} and adopted a gravitational softening length of 15 pc. The simulation includes radiative cooling over a wide temperature range ($20\,{\rm K} < T < 10^8\,{\rm K}$), heating by far-ultraviolet background radiation, probabilistic star formation from cold, dense gas, and thermal feedback from both Type II supernovae and H\,II regions \citep{Saitoh+2008,SaitohMakino2009,Baba+2017}. The initial numbers of star, gas (SPH), and dark matter particles are 5.4 million, 1 million, and 11 million, respectively, with corresponding particle masses of approximately $5 \times 10^3\,M_{\odot}$, $5 \times 10^3\,M_{\odot}$, and $5 \times 10^4\,M_{\odot}$.

To track the evolution of individual elemental abundances, we employed {\tt CELib} \citep{Saitoh2017}\footnote{https://bitbucket.org/tsaitoh/celib/src/master/}, a chemical evolution library that incorporates multiple stellar yield tables. Specifically, we adopted the Type II supernova yields of \citet{Nomoto+2013} for massive stars, and those of \citet{Karakas2010} and \citet{Doherty+2014} for low- and intermediate-mass stars. For Type Ia supernovae, we used the theoretical yields of \citet{Iwamoto+1999} (model W7), combined with a power-law delay-time distribution (DTD), $\propto t_{\rm delay}^{-1}$, within the range $t_{\rm min} \leq t_{\rm delay} \leq 10$~Gyr \citep[][]{Totani+2008}. We set $t_{\rm min} = 150$ Myr, roughly the median of observed values \citep[$t_{\rm min} \approx 40$--300 Myr;][]{Maoz+2014ARAA}. The DTD normalization, $N_{\rm Ia}$, was chosen so that the cumulative number of Type Ia SNe reaches $0.8 \times 10^{-3}\,M_{\odot}^{-1}$ by $10$~Gyr, consistent with observational estimates \citep[e.g.][]{MaozGraur2017}.

Metal mixing plays a critical role in the enrichment of galaxies. In our simulation, we account for metal diffusion by implementing a turbulent metal mixing model in DISPH \citep[][]{Saitoh2017}, based on \citet{ShenWadsley+2010}. Because our simulation cannot resolve the smallest scales of interstellar turbulence, we introduce a scaling factor for metal diffusion, $C_{\rm d} = 0.1$, as suggested by \citet{HiraiSaitoh2017}. This approach allows us to solve the diffusion equation for metals and better capture their redistribution within the interstellar medium.

We impose an initial radial profile for the mean gas metallicity as
\begin{eqnarray}
{\rm [Fe/H]_{ISM}}(R_{\rm gc}) = -0.3 - 0.03 \left(\frac{R_{\rm gc}}{1\,\rm kpc}\right),
\end{eqnarray}
where $R_{\rm gc}$ is the guiding-center radius. The slope of -0.03 dex/kpc is adopted based on classical Cepheid observations by \citet{Matsunaga+2023}. The central value of [Fe/H] = -0.3 dex is set deliberately low so that after 3.5 Gyr the simulated gas-phase metallicity approaches current observed values. At each radius, the metallicity distribution function is assumed to be Gaussian with a dispersion of 0.05 dex.

%%% バー形成期の進化
%%%%%%%%%%%%%%%%%%%
\section{Evolution around Bar Formation Epoch}
\label{sec:bar}

In this study, we follow the evolution of our simulated galaxy up to $t = 3.5\,\mathrm{Gyr}$, where the look-back time is defined as $t_{\mathrm{bk}} \equiv 3.5\,\mathrm{Gyr} - t$. A bar spontaneously forms around $t \approx 1\,\mathrm{Gyr}$ (i.e., $t_{\mathrm{bk}} \approx 2.5\,\mathrm{Gyr}$) and becomes largely stabilized by $t \approx 1.5\,\mathrm{Gyr}$ (i.e., $t_{\mathrm{bk}} \approx 2\,\mathrm{Gyr}$). 
We determine the bar’s pattern speed and length at each epoch using the method described by \citet{Dehnen+2023}\footnote{https://github.com/WalterDehnen/patternSpeed}. The bar length $R_{\rm b}$ as the geometric mean of the radius $R_{S}$ at which the bar strength
$S(R)$ attains its maximum, and the radius $R_{\Sigma RS}$ at which the quantity $\Sigma(R)\,R\,S(R)$
reaches its peak, where $\Sigma(R)$ is the the $m=0$ Fourier coefficient of the stellar surface density. The resulting the bar's length is found to be approximately 2--3 kpc.\footnote{
\citet{Hilmi+2020} advocate defining the bar length as the radius at which the relative amplitude $A_2/A_0$ falls below a given threshold. Although we have re-measured the bar length using the method proposed by \citet{Hilmi+2020}, we find that our $A_2/A_0$ profiles decline through 0.3 at approximately the same radius ($\sim 2.5$ kpc) as the geometric-mean definition adopted in this study.}
During this period, the bar's pattern speed gradually decreases from $\approx 45$~\ps{} to $\approx 38$~\ps{}, while the corotation radius ($R_{\rm CR}$) increases from roughly 4.8 to 5.8 kpc, primarily due to dynamical friction with the dark matter halo \citep[e.g.,][]{Chiba+2021}.

In our model, we define the bar age as $\tau_{\rm bar} \equiv 2.5\,\mathrm{Gyr}$, a definition chosen based on the limited duration of our simulation. It is important to note that this does not imply that the Milky Way's bar is 2.5 Gyr old; in fact, recent observational studies suggest that the Milky Way's bar is approximately 8 Gyr old \citep[e.g.,][]{Sanders+2024}. For the discussion that follows, we express our results in terms of the look-back time relative to the bar age, defined as $\tau \equiv t_{\mathrm{bk}} - \tau_{\rm bar}$.

\begin{figure*}
\begin{center}
\includegraphics[width=0.98\textwidth]{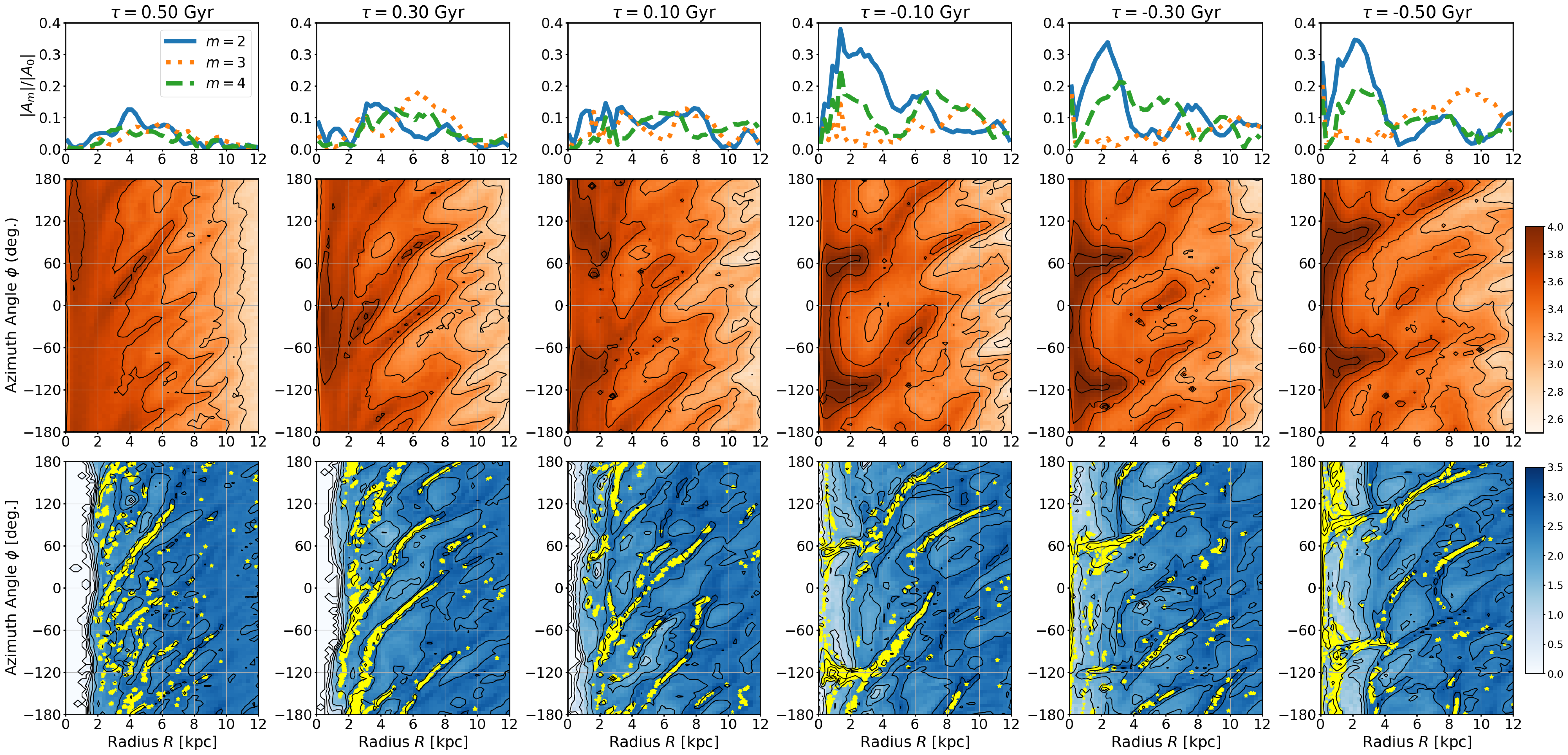}
\end{center}
\caption{
    Evolution of the simulated galaxy. The displayed time steps, from left to right, are $\tau = +0.5$, $+0.3$, $+0.1$, $-0.1$, $-0.3$, and $-0.5\,\mathrm{Gyr}$. Here, $\tau$ is defined as the look-back time ($t_{\rm bk}$) \textit{relative} to the bar age ($\tau_{\rm bar}$), i.e.\ $\tau = t_{\mathrm{bk}} - \tau_{\mathrm{bar}}$. The galaxy rotates in the $\phi < 0$ direction.
    \textbf{(Top row)} Radial profiles of the relative Fourier amplitude $|A_m|/|A_0|$ of the stellar surface density for $m=2$ (solid), 3 (dashed), and 4 (dot-dashed). 
    \textbf{(Middle row)} Stellar surface density in the simulated barred spiral galaxy, shown in polar coordinates ($R,\phi$). Colors represent the surface density (in arbitrary units) on a logarithmic scale. 
    \textbf{(Bottom row)} Same as the middle row, but for the gas surface density. Yellow star markers indicate the young star particles with an age $\le 10$ Myr.
    \textbf{Alt Text}: Six snapshots of a simulated barred spiral galaxy are shown at different times relative to bar formation. The top row displays radial Fourier amplitude profiles for stellar density modes two, three, and four. The middle row shows a polar map of stellar surface density, and the bottom row shows a polar map of gas surface density. Young stars with ages of 10 million years or less are indicated.
}
\label{fig:map}
\end{figure*}

Figure~\ref{fig:map} illustrates the evolution of the simulated galaxy during the bar formation epoch ($\tau \le 0.5$ Gyr). In the top row, we plot the radial profiles of he relative Fourier amplitude $|A_m|/|A_0|$ of the stellar surface density for modes $m=2$, 3, and 4 (depicted as solid, dashed, and dot-dashed lines, respectively), which quantify the strength and evolution of non-axisymmetric features such as the bar and spiral arms. Prior to bar formation ($\tau > 0$), the $m=2$, 3, and 4 modes clearly delineate well-developed spiral arms. Subsequently, as the bar forms, the $m=2$ mode quickly becomes dominant for $R \lesssim 4$ kpc. This trend is clearly reflected in the stellar surface density maps shown in the middle row of Figure~\ref{fig:map}. Outside the bar's end (around $R \approx 4$ kpc), the $m=2$ spiral arms remain prominent, though they do not always connect directly to the bar's end, suggesting that the bar and spiral arms rotate at different angular speeds \citep[][]{SellwoodSparke1988,Baba2015c,Fujii+2018,Hilmi+2020,Vislosky+2024}.

The bottom row of Figure~\ref{fig:map} shows the evolution of the gas surface density during the bar formation epoch. Like the stellar component, the gas distribution exhibits spiral arms whose locations match those seen in the stellar maps, but it also reveals additional, smaller-scale spiral features. Young star particles (ages $\le 10$ Myr), indicated by yellow star markers, further highlight that star formation occurs along these gas spiral arms. Notably, vigorous star formation is observed in the inner disk region ($2 \lesssim R \lesssim 6$ kpc), where the strong spiral arms developing prior to bar formation drive enhanced star formation. This evidence indicates that, prior to the full establishment of the bar, strong spiral arms in the inner disk promote active star formation.

%%% 星形成率
%%%%%%%%%%%%%%%%%%%
\section{Star Formation Segregation}
\label{sec:SFR}

\begin{figure*}
\begin{center}
\includegraphics[width=0.98\textwidth]{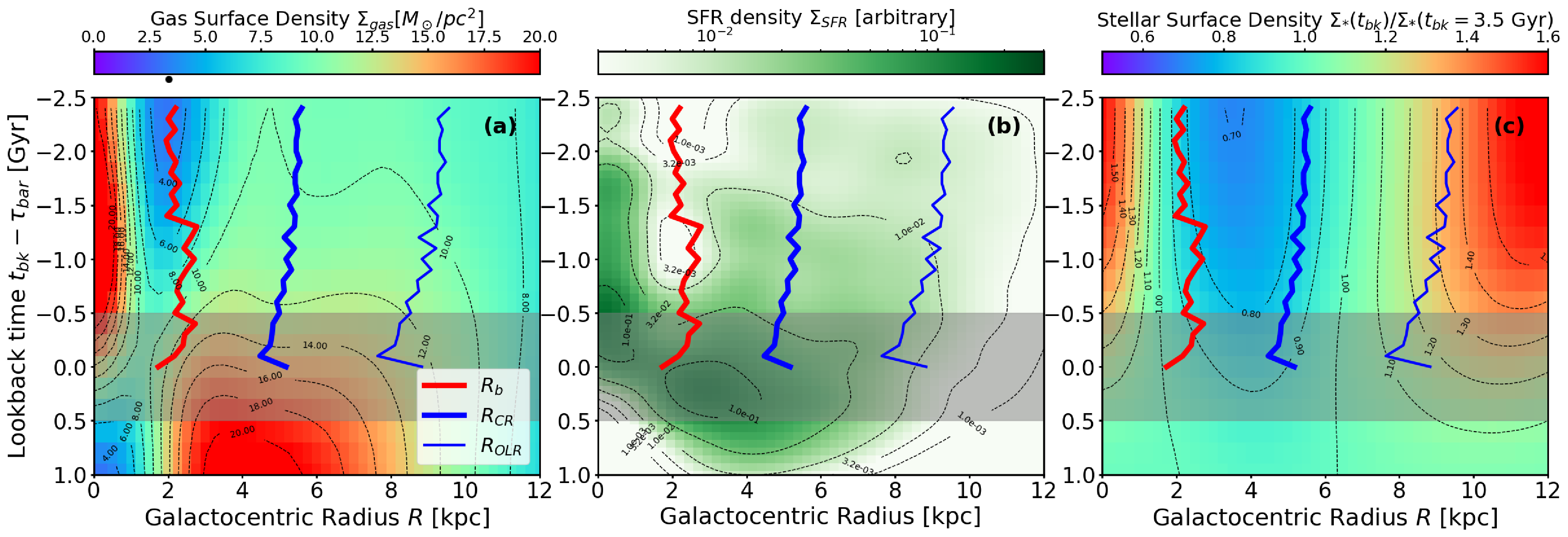}
\end{center}
\caption{
    Temporal evolution of gas, SFR, and stellar surface densities, plotted against galactocentric radius $R$ and look-back time $\tau$ $(= t_{\mathrm{bk}} - \tau_{\mathrm{bar}})$. The horizontal gray-shaded region denotes the bar formation epoch ($|\tau| < 0.5\,\mathrm{Gyr}$). Overlaid in each panel are the bar radius ($R_{\rm b}$; red solid), the corotation radius ($R_{\rm CR}$; blue solid), and the outer Lindblad resonance ($R_{\rm OLR}$; thin blue solid).  
    \textbf{(a)} Gas surface density, $\Sigma_{\rm gas}(R,\tau)$.  
    \textbf{(b)} SFR surface density, $\Sigma_{\rm SFR}(R,\tau)$.  
    \textbf{(c)} Stellar surface density ratio, $\Sigma_{\ast}(R,\tau) / \Sigma_{\ast,\rm init}(R)$, where $\Sigma_{\ast,\rm init}(R)$ is the initial stellar surface density profile.
    \textbf{Alt Text}: Three panels display the evolution of gas, star formation rate, and the ratio of current to initial stellar surface density versus galactocentric radius and look-back time relative to the bar age. The gray area indicates the bar formation epoch, and key resonant radii are overplotted.
}
\label{fig:evol}
\end{figure*}

Figure~\ref{fig:evol}(a) displays the long-term evolution of the gas surface density, $\Sigma_{\rm gas}$. Owing to the initial exponential profile with a central hole \citep[][]{McMillan2017}, $\Sigma_{\rm gas}$ peaks at around 4\,kpc. Following bar formation ($\tau \lesssim 0$), the gas surface density sharply decreases between 2 and 6\,kpc, while it markedly increases in the central region ($R \lesssim 1$\,kpc). This behavior is attributed to bar-induced gas inflows that channel gas into the central few hundred parsecs \citep[][]{Athanassoula1992a}, thereby boosting star formation within that nuclear region (see Figure~\ref{fig:evol}(b); e.g., \citealt{FriedliBenz1995,Seo+2019,BabaKawata2020a,Tress+2020}). Once the stellar bar matures ($\tau \lesssim -0.5\,\mathrm{Gyr}$), the gas in the bar region ($1 \lesssim R \lesssim 3$\,kpc) becomes depleted.

Figure~\ref{fig:evol}(b) shows that, corresponding to this gas depletion, the SFR in the bar region ($1 \lesssim R \lesssim 3$\,kpc) drops sharply after bar formation, with a similar decline observed throughout the disk \citep[e.g.][]{Spinoso+2017,George+2019}. This reduction is driven both by gas consumption via star formation and by its redistribution by the bar and spiral arms. Notably, star formation in the bar region remains active during the bar formation phase (see also the bottom row of Figure~\ref{fig:map}), but it falls precipitously once the bar is fully developed. These findings suggest that the SFR in the bar region is closely linked to the evolutionary stage of the bar \citep[][]{Verley+2007}.

To investigate the variation in star formation rate as a function of galactocentric distance, we focus on four annular regions: $2 < R < 3\,\mathrm{kpc}$, $5 < R < 6\,\mathrm{kpc}$, $8 < R < 9\,\mathrm{kpc}$, and $11 < R < 12\,\mathrm{kpc}$. These radial bins capture populations of stars formed in diverse environments, ranging from the bar-dominated inner region to the outskirts of the disk. The filled histograms (red) in the upper panels of Figure~\ref{fig:AMR_Rbirth} display the in-situ SFR in each region. These histograms were generated using Sturges’ formula to determine the bin width, and each radial bin contains tens of thousands of star particles. As described above, the SFR in the bar region ($2 < R < 3\,\mathrm{kpc}$) sharply decreases after bar formation. A similar, though somewhat more modest, decline is also observed in the inner disk outside the bar ($5 < R < 6\,\mathrm{kpc}$) and in the solar neighborhood ($8 < R < 9\,\mathrm{kpc}$). Note that in our current model, the absence of gas replenishment from the halo leads to a gradual decline in the overall gas mass over time, which further contributes to the decrease in SFR. However, our analysis shows that in the inner disk ($2 \lesssim R \lesssim 6$ kpc) most of this gas loss is driven by bar-induced inflows into the central region rather than by star formation consumption, so even with ongoing halo accretion the inner-disk SFR would not recover substantially.

In contrast, the outer disk region ($11 < R < 12\,\mathrm{kpc}$) exhibits an increasing trend in SFR after bar formation, indeed, there appears to be a rapid rise immediately following bar formation. Moreover, the in-situ SFR shows quasi-periodic fluctuations with a period of approximately 200--300 Myr, a feature that is particularly prominent in the outer disk. These results indicate that bar formation not only quenches star formation in the inner regions but also promotes a spatial segregation of star formation activity across the galactic disk. 

Interestingly, these periodic oscillations in the in-situ star formation rate--while most pronounced in the outer disk--also occur at other radii and may, for example, reflect the resonant coupling between the bar and spiral arms discussed by \citet{Marques+2025}. However, the most dramatic change in in-situ star formation coincides with the bar formation. This indicates that, even if bar–spiral resonances modulate star formation over time, the initial bar-formation event still leaves the strongest imprint on its spatial and temporal distribution.

%%%%%%%%%%%%%%%%%%%
\section{Bar-driven Migration and Age-Metallicity Distributions of Stars}
\label{sec:migration}

Observationally, the star formation history cannot be directly measured. Therefore, to study the Milky Way's star formation history, researchers analyze the frequency distribution of stellar ages \citep[][]{Mor+2019,Ruiz-Lara+2020,Sahlholdt+2022,del-Alcazar-Julia+2025}. In this study, we similarly examine the age distribution of newly formed stars. At $t = 3.5\,\mathrm{Gyr}$ (i.e., $\tau = -2.5\,\mathrm{Gyr}$), we select newly formed star particles (i.e., those that originated from gas during the simulation, excluding any particles present at $t = 0$) and analyze their age, [Fe/H], and birth radius $R_{\mathrm{birth}}$.

\begin{figure*}
\begin{center}
\includegraphics[width=0.98\textwidth]{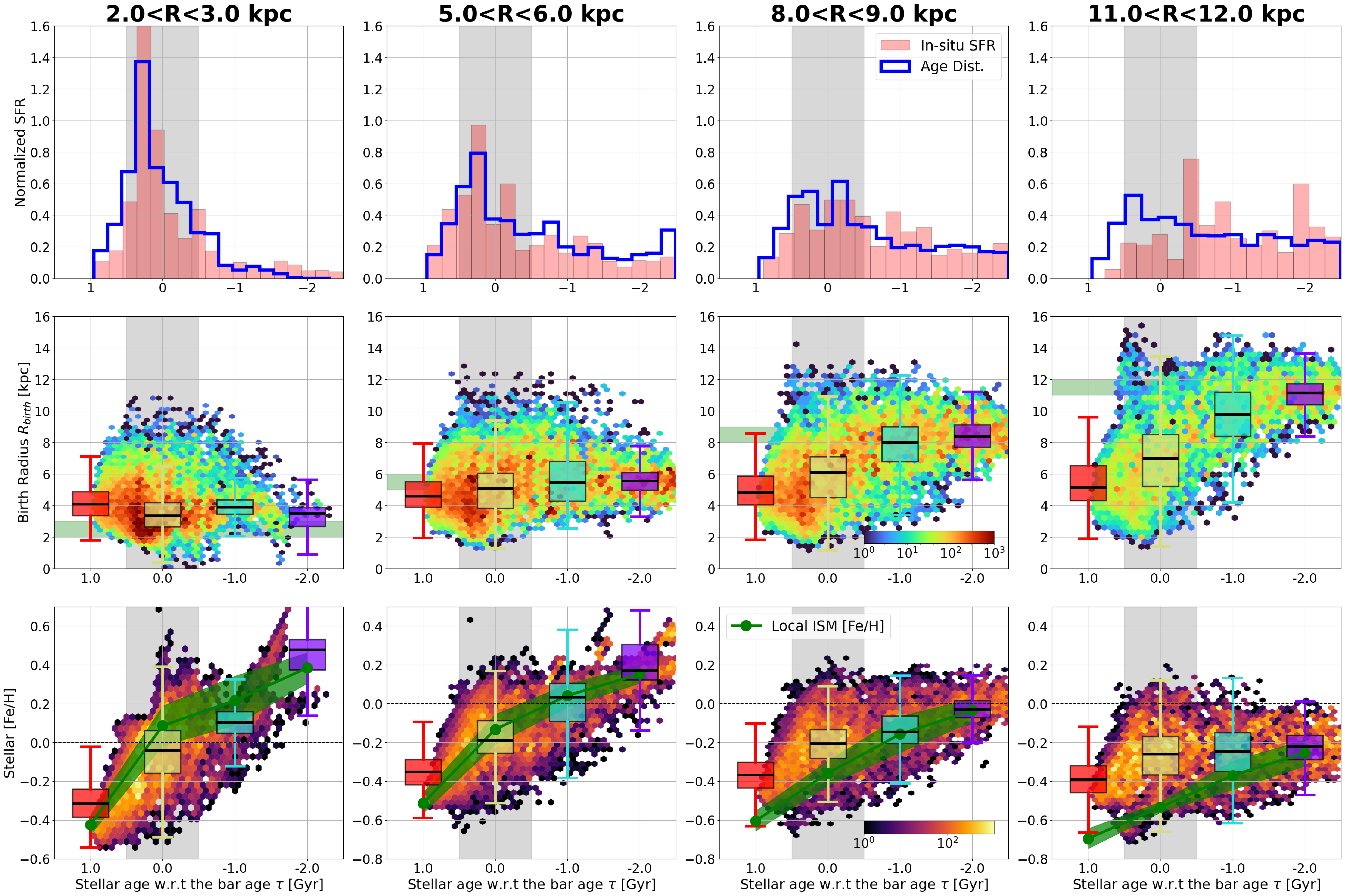}
\end{center}
\caption{
    Stellar properties at $t=3.5\,\mathrm{Gyr}$ in four galactocentric cylindrical regions: $2<R<3\,\mathrm{kpc}$ (leftmost), $5<R<6\,\mathrm{kpc}$, $8<R<9\,\mathrm{kpc}$, and $11<R<12\,\mathrm{kpc}$ (rightmost). 
    Note that the horizontal axis indicates stellar ages \textit{relative} to the bar age, $\tau$. 
    \textbf{(Top row)} Temporal evolution of In-situ SFR (filled red histogram) and age distributions of newly formed stars (blue histogram) in each cylindrical region. A pronounced peak at the relative age $\tau \approx 0$ marks the bar formation epoch (shaded in gray).
    \textbf{(Middle row)} Distribution of birth radii as a function of stellar age. The color map, shown on a logarithmic scale, represents the number density of star particles. Each panel includes box-and-whisker plots (in 1~Gyr-wide age bins) that indicate the mean and dispersion of birth radii. The results reveal substantial radial migration: stars formed before bar formation ($\tau \gtrsim 0$) tend to move outward toward the solar and outer regions or inward toward the bar-dominated zone.
    \textbf{(Bottom row)} Stellar age-metallicity relations. The color map, shown on a logarithmic scale, represents the number density of star particles. Each panel displays box-and-whisker plots of [Fe/H] for newly formed stars in 1~Gyr-wide age bins, overlaid with the mean and dispersion of the gas-phase [Fe/H] at the corresponding look-back times (green). The inner disk ($2 < R < 3\,\mathrm{kpc}$) exhibits a steeper enrichment gradient, whereas the outer regions (e.g., $R > 11\,\mathrm{kpc}$) maintain nearly constant [Fe/H] for ages younger than the bar age (i.e., $\tau \lesssim 0$). Comparing stellar [Fe/H] with the gas-phase values suggests significant radial migration, as older stars in the outer disk often have higher [Fe/H] than the local gas.
    \textbf{Alt Text}: This figure displays the stellar properties at 3.5 Gyr in four cylindrical regions of a simulated galaxy. The top row shows red histograms of the in-situ star formation rate and blue histograms of stellar ages, with a clear peak at the bar formation epoch. The middle row presents distributions of birth radii as a function of age, with a color map indicating the number density of stars, revealing significant radial migration. The bottom row shows age–metallicity relations, with the gas-phase metallicity overlaid.
}	
\label{fig:AMR_Rbirth}
\end{figure*}

The upper panels of Figure~\ref{fig:AMR_Rbirth} display the age distributions (blue histograms) in each radial region. All regions exhibit a pronounced peak at $\tau \lesssim 0$, corresponding to the bar formation epoch. This peak is most prominent in the inner disk (e.g., $2 < R < 3\,\mathrm{kpc}$) and diminishes with increasing radius. When compared with the in-situ SFR (red filled histograms), the age distribution and the in-situ SFR agree well in regions interior to the solar circle. However, in the solar neighborhood and outer disk regions, the age distribution shows a more pronounced peak at the bar formation epoch than does the in-situ SFR-a discrepancy that is especially notable in the outer disk region ($11 < R < 12\,\mathrm{kpc}$).

This discrepancy arises because, before bar formation, the inner disk experienced a period of active star formation, leading to the formation of many stars (Section \ref{sec:bar}). However, after bar formation, the SFR in the inner disk sharply declines (Figure~\ref{fig:evol}(b)), while strong spiral arms - induced during the bar formation epoch - promote vigorous star formation in the inner disk. 
Consequently, a significant fraction of stars formed in the inner disk prior to bar formation migrate outward. Figure~\ref{fig:evol}(c) shows that the stellar surface density decreases by approximately 20\% between $R_{\rm b}$ and $R_{\rm CR}$, while it increases by about 30-50\% in the central region ($R \lesssim 1\,\mathrm{kpc}$) and the outer disk ($R \gtrsim R_{\rm OLR}$). In the central region, this increase is partly due to ongoing star formation; in the outer disk, where star formation is relatively weak, the increase is primarily a result of outward migration from the inner disk region.

As a caveat, this inner disk enhancement does not arise directly from the bar but from the rapid growth of strong spiral arms. As shown in the bottom row of Figure \ref{fig:map}, these arms compress gas in the inner disk, raising its density and triggering a transient boost in star formation. In principle, any mechanism that produces such pronounced spiral structure--even in the absence of a bar--would similarly elevate the inner disk star formation rate.

The middle row of Figure~\ref{fig:AMR_Rbirth} presents a plot of stellar age versus birth radius $R_{\rm birth}$ in each cylindrical region. In regions beyond the solar circle (e.g., $R \gtrsim 8\,\mathrm{kpc}$), stars older than the bar-formation epoch ($\tau \gtrsim 0$) predominantly formed at smaller radii (approximately 4-6 kpc) and subsequently migrated outward. Conversely, in the inner region ($R < 3\,\mathrm{kpc}$), many older stars originated near $R \approx 4\,\mathrm{kpc}$ and later migrated inward, likely due to bar trapping. Notably, the mean birth radius shows a sharp transition at the bar formation epoch, providing clear evidence that stellar migration is strongly promoted during this phase \citep[][]{DiMatteo+2013,Khoperskov+2020}. Even after the bar has stabilized, continued interactions between the bar and spiral arms maintain fluctuations in the gravitational potential, driving further angular-momentum exchange and ongoing migration \citep[e.g.][]{MinchevFamaey2010,Marques+2025}.
These contrasting trends highlight the robust radial mixing associated with bar formation, whereby stars from different birth radii either converge toward or diverge from the bar-dominated region.

The impact of outward migration from the inner disk is not limited to the age distribution; it also affects the metallicity distribution. The bottom row of Figure~\ref{fig:AMR_Rbirth} presents the age-metallicity relation for the newly formed stars in each cylindrical region. For clarity, we display box-and-whisker plots that indicate the mean and dispersion of [Fe/H] in 1 Gyr-wide age bins. The bar region (e.g., $2 < R < 3\,\mathrm{kpc}$) exhibits a steeper [Fe/H] gradient with age, whereas the outermost region ($R > 11\,\mathrm{kpc}$) maintains nearly constant [Fe/H] for ages younger than the bar formation epoch ($\tau \lesssim 0$). Overlaid in green is the mean and dispersion of the gas [Fe/H] at each look-back time in the same cylindrical regions. Unlike the stellar trends, the local gas-phase metallicity increases continuously with time at all radii--most rapidly in the inner disk and more gradually in the outer disk--which matches the standard expectations of inside-out chemical evolution \citep[e.g.][]{Chiappini+2001}.

This contrast between the steadily rising gas‐phase [Fe/H] and the flat stellar [Fe/H] in the solar circle and outer disk reflects the well-established process of radial migration \citep[e.g.][]{Hayden+2015,Feuillet+2018,Dantas+2022,Lehmann+2024}. In our simulation, this migration is particularly pronounced around the bar-formation epoch: many of the stars in the outer disk that are more metal-rich than the local ISM at their current radius are older than the bar's age, indicating that they were born in the metal-rich inner regions and migrated outward.

It is worth noting that the stellar [Fe/H] dispersions in Figure \ref{fig:AMR_Rbirth} (bottom row) are substantially larger (about 0.3–0.4 dex) than the approximately 0.05 dex dispersion of the local interstellar medium [Fe/H] (green bands). This larger spread arises because stars in each radial bin originate from a wide range of birth radii (see also the middle row of Figure \ref{fig:AMR_Rbirth}), reflecting a mixture of metal-rich inner disk and metal-poor outer disk populations. This behavior closely aligns with the age-metallicity relation reported by \citet{Lu+arXiv221204515}, in which the upper boundary of the age-metallicity relation is defined by inner disk stars and the lower boundary by outer disk stars.

%%%%%%%%%%%%%%%%%%%
\section{Discussion}
\label{sec:Discussion}

Our results indicate that a substantial fraction of stars formed in the inner disk before bar formation later migrate outward. This effect is driven by two key processes. First, once the bar forms, star formation in the inner disk is suppressed, a phenomenon often referred to as ``bar quenching'' \citep[e.g.,][]{Spinoso+2017}, which results in a lower production rate of stars younger than the bar age. Second, during the bar formation phase, strong spiral arms develop that not only boost star formation in the inner disk but also enhance the outward transport of angular momentum \citep[e.g.,][]{DiMatteo+2013}. In combination, these processes lead to an increased number of stars with ages corresponding to the bar formation epoch and with metallicities characteristic of the inner disk appearing in the outer regions.

More broadly, while our simulation links the pre-bar burst to spiral-arm growth, in real galaxies early inner disk activity may be driven by cosmological gas inflows \citep[e.g.][]{Chiappini+1997}, dissipative collapse, or other dynamical processes \citep[e.g.][]{Brook+2004}. What fundamentally shapes the later evolution is the segregation of star formation activities induced by bar formation (Section 4). Once the bar is established, it quenches inner disk star formation while allowing outer-disk formation to continue and drives the outward migration of older inner disk stars. This bar-driven segregation imprints characteristic signatures on the age and metallicity distributions of the disk, independently of the physical origin of the initial star formation enhancement.

\citet{Sanders+2024} applied the methods of \citet{BabaKawata2020a} and \citet{Baba+2022} to the age distribution of Mira variables in the nuclear stellar disk and the outer boxy or peanut-shaped bulge, suggesting that the Milky Way's bar is $8 \pm 1$ Gyr old. Our simulation shows that the bar formation epoch produces a pronounced peak in the disk-star age distribution. Thus, if the Milky Way's bar did indeed form around 8 Gyr ago \citep[][]{Bovy+2019,Nogueras-Lara+2020,Wylie+2022,Schodel+2023,Haywood+2024,Sanders+2024}, as these observations imply, one would expect to observe a corresponding peak in the age distribution of disk stars.

Recent observational studies have reported an increase in the stellar age distribution at around 8 to 9 Gyr \citep[][]{Snaith+2015,Mor+2019,del-Alcazar-Julia+2025}. However, \citet{Sahlholdt+2022} did not detect a pronounced 8 Gyr peak in the age distribution of outer disk stars, and \citet{Ruiz-Lara+2020} analyzed the age distribution within 2 kpc of the solar neighborhood, reporting peaks at ages corresponding to the pericentric passages of the Sagittarius dwarf galaxy (approximately 1 Gyr, 1.7 Gyr, and 5 Gyr). Although observational studies consistently report variability in the stellar age distribution, there is currently no consensus regarding the precise location of these peaks \citep[][]{Snaith+2015,Mor+2019,Ruiz-Lara+2020,Sahlholdt+2022,del-Alcazar-Julia+2025}. Therefore, the peak in the stellar age distribution associated with the bar formation epoch, as suggested by our study, does not entirely match current observational findings.

This discrepancy likely arises because our model does not fully capture the complex evolution of the Milky Way. For example, although our current model does not exhibit strong bar slowdown, extended simulations may reveal additional migration effects associated with bar deceleration \citep[e.g.,][]{Halle+2015,Khoperskov+2020,Chiba+2021,Baba+2024}. Recently, \citet{Zhang+2025} noted that, due to bar deceleration, old and metal-rich stars from the inner disk can become trapped in the bar corotation resonance and migrate together to around 6 to 7 kpc, resulting in two distinct sequences in the stellar age-metallicity distribution. Moreover, quasi-periodic perturbations from the Sagittarius dwarf galaxy may also affect the star formation activity \citep[][]{Ruiz-Lara+2020,AnnemKhoperskov2022} and the efficiency of stellar migration \citep[e.g.][]{Bird+2012,TsujimotoBaba2019,Carr+2022} in the Galactic disk. Combined, these additional effects may lead to a more complex age-metallicity structure than that indicated by our present study.

It is important to emphasize that our study is not primarily aimed at reproducing all current observational results. Instead, our focus is on elucidating the impact of bar formation on the evolution of the stellar disk. 
Our simulation shows that the combination of inner disk star formation--enhanced by strong spiral arms or other dynamical processes prior to bar formation and subsequently suppressed once the bar is established--together with the bar-driven stellar migration, can produce a distinctive peak in the outer disk's age distribution. We also note that resonant interactions between the bar and spiral arms may further modulate star formation and stellar migration over time \citep[][]{MinchevFamaey2010,Marques+2025}.

Our findings offer a conceptual framework for understanding the long-term influence of bar-induced secular evolution on the Galactic disk. Moreover, given that observational studies report significant variability in the stellar age distribution and that the precise locations of these peaks remain a matter of debate, caution is warranted when interpreting the stellar age distribution as a direct measure of the star formation rate. In future work, we plan to extend our simulations to incorporate additional dynamical processes, such as interactions with the Sagittarius dwarf galaxy \citep[e.g.][]{Laporte+2019,Hunt+2021,Asano+2025}, enabling a more comprehensive comparison with observations.

\begin{ack}
We thank the anonymous referee for their constructive comments and suggestions, which helped improve the clarity and quality of the manuscript. We are grateful to Takayuki Saitoh for providing and supporting the {\tt ASURA-3} simulation code, and to Takuji Tsujimoto for valuable discussions regarding the interpretation of our results. 
Calculations, numerical analyses and visualization were carried out on Cray XC50 (ATERUI-II) and computers at the Center for Computational Astrophysics, National Astronomical Observatory of Japan (CfCA/NAOJ). This research was supported by the Japan Society for the Promotion of Science (JSPS) under Grant Numbers 21K03633, 21H00054, 22H01259, 24K07095, and 25H00394.
\end{ack}

%%%
%\bibliographystyle{mn2e}
%\bibliography{ms}

\begin{thebibliography}{88}
\expandafter\ifx\csname natexlab\endcsname\relax\def\natexlab#1{#1}\fi

\bibitem[{{Anders} {et~al}\mbox{.}(2019){Anders}, {Khalatyan}, {Chiappini},
  {Queiroz}, {Santiago}, {Jordi}, {Girardi}, {Brown}, {Matijevi{\v{c}}},
  {Monari}, {Cantat-Gaudin}, {Weiler}, {Khan}, {Miglio}, {Carrillo},
  {Romero-G{\'o}mez}, {Minchev}, {de Jong}, {Antoja}, {Ramos}, {Steinmetz}, \&
  {Enke}}]{Anders+2019}
{Anders} F. {et~al.}, 2019, \aap, 628, A94

\bibitem[{{Annem} \& {Khoperskov}(2024)}]{AnnemKhoperskov2022}
{Annem} B., {Khoperskov} S., 2024, \mnras, 527, 2426

\bibitem[{{Asano} {et~al}\mbox{.}(2025){Asano}, {Fujii}, {Baba}, {Portegies
  Zwart}, \& {B{\'e}dorf}}]{Asano+2025}
{Asano} T., {Fujii} M.~S., {Baba} J., {Portegies Zwart} S., {B{\'e}dorf} J.,
  2025, arXiv e-prints, arXiv:2501.12436

\bibitem[{{Athanassoula}(1992)}]{Athanassoula1992a}
{Athanassoula} E., 1992, MNRAS, 259, 328

\bibitem[{{Baba}(2015)}]{Baba2015c}
{Baba} J., 2015, \mnras, 454, 2954

\bibitem[{{Baba} \& {Kawata}(2020)}]{BabaKawata2020a}
{Baba} J., {Kawata} D., 2020, \mnras, 492, 4500

\bibitem[{{Baba}, {Kawata} \& {Sch{\"o}nrich}(2022){Baba}, {Kawata}, \&
  {Sch{\"o}nrich}}]{Baba+2022}
{Baba} J., {Kawata} D., {Sch{\"o}nrich} R., 2022, \mnras, 513, 2850

\bibitem[{{Baba}, {Morokuma-Matsui} \& {Saitoh}(2017){Baba}, {Morokuma-Matsui},
  \& {Saitoh}}]{Baba+2017}
{Baba} J., {Morokuma-Matsui} K., {Saitoh} T.~R., 2017, \mnras, 464, 246

\bibitem[{{Baba}, {Tsujimoto} \& {Saitoh}(2024){Baba}, {Tsujimoto}, \&
  {Saitoh}}]{Baba+2024}
{Baba} J., {Tsujimoto} T., {Saitoh} T.~R., 2024, \apjl, 976, L29

\bibitem[{{Bird}, {Kazantzidis} \& {Weinberg}(2012){Bird}, {Kazantzidis}, \&
  {Weinberg}}]{Bird+2012}
{Bird} J.~C., {Kazantzidis} S., {Weinberg} D.~H., 2012, \mnras, 420, 913

\bibitem[{{Bland-Hawthorn} \& {Gerhard}(2016)}]{Bland-HawthornGerhard2016}
{Bland-Hawthorn} J., {Gerhard} O., 2016, \araa, 54, 529

\bibitem[{{Blitz} \& {Spergel}(1991)}]{BlitzSpergel1991a}
{Blitz} L., {Spergel} D.~N., 1991, \apj, 370, 205

\bibitem[{{Bovy} {et~al}\mbox{.}(2019){Bovy}, {Leung}, {Hunt}, {Mackereth},
  {Garc{\'\i}a-Hern{\'a}ndez}, \& {Roman-Lopes}}]{Bovy+2019}
{Bovy} J., {Leung} H.~W., {Hunt} J. A.~S., {Mackereth} J.~T.,
  {Garc{\'\i}a-Hern{\'a}ndez} D.~A., {Roman-Lopes} A., 2019, \mnras, 490, 4740

\bibitem[{{Brook} {et~al}\mbox{.}(2004){Brook}, {Kawata}, {Gibson}, \&
  {Freeman}}]{Brook+2004}
{Brook} C.~B., {Kawata} D., {Gibson} B.~K., {Freeman} K.~C., 2004, \apj, 612,
  894

\bibitem[{{Carr} {et~al}\mbox{.}(2022){Carr}, {Johnston}, {Laporte}, \&
  {Ness}}]{Carr+2022}
{Carr} C., {Johnston} K.~V., {Laporte} C. F.~P., {Ness} M.~K., 2022, \mnras,
  516, 5067

\bibitem[{{Chiappini}, {Matteucci} \& {Gratton}(1997){Chiappini}, {Matteucci},
  \& {Gratton}}]{Chiappini+1997}
{Chiappini} C., {Matteucci} F., {Gratton} R., 1997, \apj, 477, 765

\bibitem[{{Chiappini}, {Matteucci} \& {Romano}(2001){Chiappini}, {Matteucci},
  \& {Romano}}]{Chiappini+2001}
{Chiappini} C., {Matteucci} F., {Romano} D., 2001, \apj, 554, 1044

\bibitem[{{Chiba}, {Friske} \& {Sch{\"o}nrich}(2021){Chiba}, {Friske}, \&
  {Sch{\"o}nrich}}]{Chiba+2021}
{Chiba} R., {Friske} J. K.~S., {Sch{\"o}nrich} R., 2021, \mnras, 500, 4710

\bibitem[{{Dantas} {et~al}\mbox{.}(2023){Dantas}, {Smiljanic}, {Boesso},
  {Rocha-Pinto}, {Magrini}, {Guiglion}, {Tautvai{\v{s}}ien{\.{e}}}, {Gilmore},
  {Randich}, {Bensby}, {Bragaglia}, {Bergemann}, {Carraro}, {Jofr{\'e}}, \&
  {Zaggia}}]{Dantas+2022}
{Dantas} M.~L.~L. {et~al.}, 2023, \aap, 669, A96

\bibitem[{{Dehnen}, {Semczuk} \& {Sch{\"o}nrich}(2023){Dehnen}, {Semczuk}, \&
  {Sch{\"o}nrich}}]{Dehnen+2023}
{Dehnen} W., {Semczuk} M., {Sch{\"o}nrich} R., 2023, \mnras, 518, 2712

\bibitem[{{del Alc{\'a}zar-Juli{\`a}} {et~al}\mbox{.}(2025){del
  Alc{\'a}zar-Juli{\`a}}, {Figueras}, {Robin}, {Bienaym{\'e}}, \&
  {Anders}}]{del-Alcazar-Julia+2025}
{del Alc{\'a}zar-Juli{\`a}} M., {Figueras} F., {Robin} A.~C., {Bienaym{\'e}}
  O., {Anders} F., 2025, arXiv e-prints, arXiv:2501.17236

\bibitem[{{Di Matteo} {et~al}\mbox{.}(2013){Di Matteo}, {Haywood}, {Combes},
  {Semelin}, \& {Snaith}}]{DiMatteo+2013}
{Di Matteo} P., {Haywood} M., {Combes} F., {Semelin} B., {Snaith} O.~N., 2013,
  \aap, 553, A102

\bibitem[{{Doherty} {et~al}\mbox{.}(2014){Doherty}, {Gil-Pons}, {Lau},
  {Lattanzio}, \& {Siess}}]{Doherty+2014}
{Doherty} C.~L., {Gil-Pons} P., {Lau} H. H.~B., {Lattanzio} J.~C., {Siess} L.,
  2014, \mnras, 437, 195

\bibitem[{{Feuillet} {et~al}\mbox{.}(2018){Feuillet}, {Bovy}, {Holtzman},
  {Weinberg}, {Garc{\'\i}a-Hern{\'a}ndez}, {Hearty}, {Majewski}, {Roman-Lopes},
  {Rybizki}, \& {Zamora}}]{Feuillet+2018}
{Feuillet} D.~K. {et~al.}, 2018, \mnras, 477, 2326

\bibitem[{{Fragkoudi} {et~al}\mbox{.}(2020){Fragkoudi}, {Grand}, {Pakmor},
  {Bl{\'a}zquez-Calero}, {Gargiulo}, {Gomez}, {Marinacci}, {Monachesi}, {Ness},
  {Perez}, {Tissera}, \& {White}}]{Fragkoudi+2020}
{Fragkoudi} F. {et~al.}, 2020, \mnras, 494, 5936

\bibitem[{{Friedli} \& {Benz}(1995)}]{FriedliBenz1995}
{Friedli} D., {Benz} W., 1995, \aap, 301, 649

\bibitem[{{Friedli}, {Benz} \& {Kennicutt}(1994){Friedli}, {Benz}, \&
  {Kennicutt}}]{Friedli+1994}
{Friedli} D., {Benz} W., {Kennicutt} R., 1994, \apjl, 430, L105

\bibitem[{{Fujii} {et~al}\mbox{.}(2018){Fujii}, {B{\'e}dorf}, {Baba}, \&
  {Portegies Zwart}}]{Fujii+2018}
{Fujii} M.~S., {B{\'e}dorf} J., {Baba} J., {Portegies Zwart} S., 2018, \mnras,
  477, 1451

\bibitem[{{George}, {Subramanian} \& {Paul}(2019){George}, {Subramanian}, \&
  {Paul}}]{George+2019}
{George} K., {Subramanian} S., {Paul} K.~T., 2019, \aap, 628, A24

\bibitem[{{Grand}, {Kawata} \& {Cropper}(2012){Grand}, {Kawata}, \&
  {Cropper}}]{Grand+2012b}
{Grand} R.~J.~J., {Kawata} D., {Cropper} M., 2012, MNRAS, 426, 167

\bibitem[{{Grand}, {Kawata} \& {Cropper}(2015){Grand}, {Kawata}, \&
  {Cropper}}]{Grand+2015}
{Grand} R. J.~J., {Kawata} D., {Cropper} M., 2015, \mnras, 447, 4018

\bibitem[{{Halle} {et~al}\mbox{.}(2015){Halle}, {Di Matteo}, {Haywood}, \&
  {Combes}}]{Halle+2015}
{Halle} A., {Di Matteo} P., {Haywood} M., {Combes} F., 2015, \aap, 578, A58

\bibitem[{{Hayden} {et~al}\mbox{.}(2015){Hayden}, {Bovy}, {Holtzman},
  {Nidever}, {Bird}, {Weinberg}, {Andrews}, {Majewski}, {Allende Prieto},
  {Anders}, {Beers}, {Bizyaev}, {Chiappini}, {Cunha}, {Frinchaboy},
  {Garc{\'\i}a-Her{\'n}and ez}, {Garc{\'\i}a P{\'e}rez}, {Girardi}, {Harding},
  {Hearty}, {Johnson}, {M{\'e}sz{\'a}ros}, {Minchev}, {O'Connell}, {Pan},
  {Robin}, {Schiavon}, {Schneider}, {Schultheis}, {Shetrone}, {Skrutskie},
  {Steinmetz}, {Smith}, {Wilson}, {Zamora}, \& {Zasowski}}]{Hayden+2015}
{Hayden} M.~R. {et~al.}, 2015, \apj, 808, 132

\bibitem[{{Haywood} {et~al}\mbox{.}(2024){Haywood}, {Khoperskov}, {Cerqui}, {Di
  Matteo}, {Katz}, \& {Snaith}}]{Haywood+2024}
{Haywood} M., {Khoperskov} S., {Cerqui} V., {Di Matteo} P., {Katz} D., {Snaith}
  O., 2024, \aap, 690, A147

\bibitem[{{Haywood} {et~al}\mbox{.}(2016){Haywood}, {Lehnert}, {Di Matteo},
  {Snaith}, {Schultheis}, {Katz}, \& {G{\'o}mez}}]{Haywood+2016a}
{Haywood} M., {Lehnert} M.~D., {Di Matteo} P., {Snaith} O., {Schultheis} M.,
  {Katz} D., {G{\'o}mez} A., 2016, \aap, 589, A66

\bibitem[{{Hilmi} {et~al}\mbox{.}(2020){Hilmi}, {Minchev}, {Buck}, {Martig},
  {Quillen}, {Monari}, {Famaey}, {de Jong}, {Laporte}, {Read}, {Sand ers},
  {Steinmetz}, \& {Wegg}}]{Hilmi+2020}
{Hilmi} T. {et~al.}, 2020, \mnras, 497, 933

\bibitem[{{Hirai} \& {Saitoh}(2017)}]{HiraiSaitoh2017}
{Hirai} Y., {Saitoh} T.~R., 2017, \apjl, 838, L23

\bibitem[{{Hunt} {et~al}\mbox{.}(2021){Hunt}, {Stelea}, {Johnston}, {Gandhi},
  {Laporte}, \& {B{\'e}dorf}}]{Hunt+2021}
{Hunt} J. A.~S., {Stelea} I.~A., {Johnston} K.~V., {Gandhi} S.~S., {Laporte} C.
  F.~P., {B{\'e}dorf} J., 2021, \mnras, 508, 1459

\bibitem[{{Hunt} \& {Vasiliev}(2025)}]{HuntVasiliev2025}
{Hunt} J. A.~S., {Vasiliev} E., 2025, \nar, 100, 101721

\bibitem[{{Iwamoto} {et~al}\mbox{.}(1999){Iwamoto}, {Brachwitz}, {Nomoto},
  {Kishimoto}, {Umeda}, {Hix}, \& {Thielemann}}]{Iwamoto+1999}
{Iwamoto} K., {Brachwitz} F., {Nomoto} K., {Kishimoto} N., {Umeda} H., {Hix}
  W.~R., {Thielemann} F.-K., 1999, \apjs, 125, 439

\bibitem[{{Karakas}(2010)}]{Karakas2010}
{Karakas} A.~I., 2010, \mnras, 403, 1413

\bibitem[{{Khoperskov} {et~al}\mbox{.}(2020){Khoperskov}, {Di Matteo},
  {Haywood}, {G{\'o}mez}, \& {Snaith}}]{Khoperskov+2020}
{Khoperskov} S., {Di Matteo} P., {Haywood} M., {G{\'o}mez} A., {Snaith} O.~N.,
  2020, \aap, 638, A144

\bibitem[{{Kubryk}, {Prantzos} \& {Athanassoula}(2013){Kubryk}, {Prantzos}, \&
  {Athanassoula}}]{Kubryk+2013}
{Kubryk} M., {Prantzos} N., {Athanassoula} E., 2013, \mnras, 436, 1479

\bibitem[{{Laporte} {et~al}\mbox{.}(2019){Laporte}, {Minchev}, {Johnston}, \&
  {G{\'o}mez}}]{Laporte+2019}
{Laporte} C. F.~P., {Minchev} I., {Johnston} K.~V., {G{\'o}mez} F.~A., 2019,
  \mnras, 485, 3134

\bibitem[{{Lehmann} {et~al}\mbox{.}(2024){Lehmann}, {Feltzing}, {Feuillet}, \&
  {Kordopatis}}]{Lehmann+2024}
{Lehmann} C., {Feltzing} S., {Feuillet} D., {Kordopatis} G., 2024, \mnras, 533,
  538

\bibitem[{{Li}, {Shen} \& {Kim}(2015){Li}, {Shen}, \& {Kim}}]{Li+2015}
{Li} Z., {Shen} J., {Kim} W.-T., 2015, \apj, 806, 150

\bibitem[{{Lu} {et~al}\mbox{.}(2024){Lu}, {Minchev}, {Buck}, {Khoperskov},
  {Steinmetz}, {Libeskind}, {Cescutti}, {Freeman}, \& {Ratcliffe}}]{Lu+2024}
{Lu} Y.~L. {et~al.}, 2024, \mnras, 535, 392

\bibitem[{{Maoz} \& {Graur}(2017)}]{MaozGraur2017}
{Maoz} D., {Graur} O., 2017, \apj, 848, 25

\bibitem[{{Maoz}, {Mannucci} \& {Nelemans}(2014){Maoz}, {Mannucci}, \&
  {Nelemans}}]{Maoz+2014ARAA}
{Maoz} D., {Mannucci} F., {Nelemans} G., 2014, \araa, 52, 107

\bibitem[{{Marques} {et~al}\mbox{.}(2025){Marques}, {Minchev}, {Ratcliffe},
  {Khoperskov}, {Steinmetz}, {Wenger}, {Buck}, {Martig}, {Kordopatis},
  {Schultheis}, \& {Zucker}}]{Marques+2025}
{Marques} L. {et~al.}, 2025, arXiv e-prints, arXiv:2502.02651

\bibitem[{{Matsunaga} {et~al}\mbox{.}(2023){Matsunaga}, {Taniguchi}, {Elgueta},
  {Tsujimoto}, {Baba}, {McWilliam}, {Otsubo}, {Sarugaku}, {Takeuchi}, {Katoh},
  {Hamano}, {Ikeda}, {Kawakita}, {Hull}, {Albarrac{\'\i}n}, {Bono}, \&
  {D'Orazi}}]{Matsunaga+2023}
{Matsunaga} N. {et~al.}, 2023, \apj, 954, 198

\bibitem[{{McMillan}(2017)}]{McMillan2017}
{McMillan} P.~J., 2017, \mnras, 465, 76

\bibitem[{{Minchev}, {Chiappini} \& {Martig}(2014){Minchev}, {Chiappini}, \&
  {Martig}}]{Minchev+2014b}
{Minchev} I., {Chiappini} C., {Martig} M., 2014, \aap, 572, A92

\bibitem[{{Minchev} \& {Famaey}(2010)}]{MinchevFamaey2010}
{Minchev} I., {Famaey} B., 2010, \apj, 722, 112

\bibitem[{{Mor} {et~al}\mbox{.}(2019){Mor}, {Robin}, {Figueras},
  {Roca-F{\`a}brega}, \& {Luri}}]{Mor+2019}
{Mor} R., {Robin} A.~C., {Figueras} F., {Roca-F{\`a}brega} S., {Luri} X., 2019,
  \aap, 624, L1

\bibitem[{{Nakada} {et~al}\mbox{.}(1991){Nakada}, {Onaka}, {Yamamura},
  {Deguchi}, {Hashimoto}, {Izumiura}, \& {Sekiguchi}}]{Nakada+1991}
{Nakada} Y., {Onaka} T., {Yamamura} I., {Deguchi} S., {Hashimoto} O.,
  {Izumiura} H., {Sekiguchi} K., 1991, \nat, 353, 140

\bibitem[{{Nepal} {et~al}\mbox{.}(2024){Nepal}, {Chiappini}, {Guiglion},
  {Steinmetz}, {P{\'e}rez-Villegas}, {Queiroz}, {Miglio}, {Dohme}, \&
  {Khalatyan}}]{Nepal+2024}
{Nepal} S. {et~al.}, 2024, \aap, 681, L8

\bibitem[{{Nogueras-Lara} {et~al}\mbox{.}(2020){Nogueras-Lara}, {Sch{\"o}del},
  {Gallego-Calvente}, {Gallego-Cano}, {Shahzamanian}, {Dong}, {Neumayer},
  {Hilker}, {Najarro}, {Nishiyama}, {Feldmeier-Krause}, {Girard}, \&
  {Cassisi}}]{Nogueras-Lara+2020}
{Nogueras-Lara} F. {et~al.}, 2020, Nature Astronomy, 4, 377

\bibitem[{{Nomoto}, {Kobayashi} \& {Tominaga}(2013){Nomoto}, {Kobayashi}, \&
  {Tominaga}}]{Nomoto+2013}
{Nomoto} K., {Kobayashi} C., {Tominaga} N., 2013, \araa, 51, 457

\bibitem[{{Ro{\v{s}}kar} {et~al}\mbox{.}(2008){Ro{\v{s}}kar}, {Debattista},
  {Stinson}, {Quinn}, {Kaufmann}, \& {Wadsley}}]{Roskar+2008a}
{Ro{\v{s}}kar} R., {Debattista} V.~P., {Stinson} G.~S., {Quinn} T.~R.,
  {Kaufmann} T., {Wadsley} J., 2008, \apjl, 675, L65

\bibitem[{{Ruiz-Lara} {et~al}\mbox{.}(2020){Ruiz-Lara}, {Gallart}, {Bernard},
  \& {Cassisi}}]{Ruiz-Lara+2020}
{Ruiz-Lara} T., {Gallart} C., {Bernard} E.~J., {Cassisi} S., 2020, Nature
  Astronomy, 4, 965

\bibitem[{{Sahlholdt}, {Feltzing} \& {Feuillet}(2022){Sahlholdt}, {Feltzing},
  \& {Feuillet}}]{Sahlholdt+2022}
{Sahlholdt} C.~L., {Feltzing} S., {Feuillet} D.~K., 2022, \mnras, 510, 4669

\bibitem[{{Saitoh}(2017)}]{Saitoh2017}
{Saitoh} T.~R., 2017, \aj, 153, 85

\bibitem[{{Saitoh} {et~al}\mbox{.}(2008){Saitoh}, {Daisaka}, {Kokubo},
  {Makino}, {Okamoto}, {Tomisaka}, {Wada}, \& {Yoshida}}]{Saitoh+2008}
{Saitoh} T.~R., {Daisaka} H., {Kokubo} E., {Makino} J., {Okamoto} T.,
  {Tomisaka} K., {Wada} K., {Yoshida} N., 2008, \pasj, 60, 667

\bibitem[{{Saitoh} \& {Makino}(2009)}]{SaitohMakino2009}
{Saitoh} T.~R., {Makino} J., 2009, \apjl, 697, L99

\bibitem[{{Saitoh} \& {Makino}(2013)}]{SaitohMakino2013}
{Saitoh} T.~R., {Makino} J., 2013, \apj, 768, 44

\bibitem[{{Sanders} {et~al}\mbox{.}(2024){Sanders}, {Kawata}, {Matsunaga},
  {Sormani}, {Smith}, {Minniti}, \& {Gerhard}}]{Sanders+2024}
{Sanders} J.~L., {Kawata} D., {Matsunaga} N., {Sormani} M.~C., {Smith} L.~C.,
  {Minniti} D., {Gerhard} O., 2024, \mnras, 530, 2972

\bibitem[{{Sch{\"o}del} {et~al}\mbox{.}(2023){Sch{\"o}del}, {Nogueras-Lara},
  {Hosek}, {Do}, {Lu}, {Mart{\'\i}nez Arranz}, {Ghez}, {Rich}, {Gardini},
  {Gallego-Cano}, {Cano Gonz{\'a}lez}, \& {Gallego-Calvente}}]{Schodel+2023}
{Sch{\"o}del} R. {et~al.}, 2023, \aap, 672, L8

\bibitem[{{Sellwood}(2014)}]{Sellwood2014reviw}
{Sellwood} J.~A., 2014, Reviews of Modern Physics, 86, 1

\bibitem[{{Sellwood} \& {Binney}(2002)}]{SellwoodBinney2002}
{Sellwood} J.~A., {Binney} J.~J., 2002, MNRAS, 336, 785

\bibitem[{{Sellwood} \& {Sparke}(1988)}]{SellwoodSparke1988}
{Sellwood} J.~A., {Sparke} L.~S., 1988, MNRAS, 231, 25P

\bibitem[{{Seo} {et~al}\mbox{.}(2019){Seo}, {Kim}, {Kwak}, {Hsieh}, {Han}, \&
  {Hopkins}}]{Seo+2019}
{Seo} W.-Y., {Kim} W.-T., {Kwak} S., {Hsieh} P.-Y., {Han} C., {Hopkins} P.~F.,
  2019, \apj, 872, 5

\bibitem[{{Shen}, {Wadsley} \& {Stinson}(2010){Shen}, {Wadsley}, \&
  {Stinson}}]{ShenWadsley+2010}
{Shen} S., {Wadsley} J., {Stinson} G., 2010, \mnras, 407, 1581

\bibitem[{{Snaith} {et~al}\mbox{.}(2015){Snaith}, {Haywood}, {Di Matteo},
  {Lehnert}, {Combes}, {Katz}, \& {G{\'o}mez}}]{Snaith+2015}
{Snaith} O., {Haywood} M., {Di Matteo} P., {Lehnert} M.~D., {Combes} F., {Katz}
  D., {G{\'o}mez} A., 2015, \aap, 578, A87

\bibitem[{{Sormani} {et~al}\mbox{.}(2020){Sormani}, {Tress}, {Glover},
  {Klessen}, {Battersby}, {Clark}, {Hatchfield}, \& {Smith}}]{Sormani+2020}
{Sormani} M.~C., {Tress} R.~G., {Glover} S. C.~O., {Klessen} R.~S., {Battersby}
  C.~D., {Clark} P.~C., {Hatchfield} H.~P., {Smith} R.~J., 2020, \mnras, 497,
  5024

\bibitem[{{Sormani} {et~al}\mbox{.}(2018){Sormani}, {Tre{\ss}}, {Ridley},
  {Glover}, {Klessen}, {Binney}, {Magorrian}, \& {Smith}}]{Sormani+2018a}
{Sormani} M.~C., {Tre{\ss}} R.~G., {Ridley} M., {Glover} S. C.~O., {Klessen}
  R.~S., {Binney} J., {Magorrian} J., {Smith} R., 2018, \mnras, 475, 2383

\bibitem[{{Spinoso} {et~al}\mbox{.}(2017){Spinoso}, {Bonoli}, {Dotti}, {Mayer},
  {Madau}, \& {Bellovary}}]{Spinoso+2017}
{Spinoso} D., {Bonoli} S., {Dotti} M., {Mayer} L., {Madau} P., {Bellovary} J.,
  2017, \mnras, 465, 3729

\bibitem[{{Tanikawa} {et~al}\mbox{.}(2013){Tanikawa}, {Yoshikawa}, {Nitadori},
  \& {Okamoto}}]{Tanikawa+2013}
{Tanikawa} A., {Yoshikawa} K., {Nitadori} K., {Okamoto} T., 2013, New A., 19,
  74

\bibitem[{{Totani} {et~al}\mbox{.}(2008){Totani}, {Morokuma}, {Oda}, {Doi}, \&
  {Yasuda}}]{Totani+2008}
{Totani} T., {Morokuma} T., {Oda} T., {Doi} M., {Yasuda} N., 2008, \pasj, 60,
  1327

\bibitem[{{Tress} {et~al}\mbox{.}(2020){Tress}, {Sormani}, {Glover}, {Klessen},
  {Battersby}, {Clark}, {Hatchfield}, \& {Smith}}]{Tress+2020}
{Tress} R.~G., {Sormani} M.~C., {Glover} S. C.~O., {Klessen} R.~S., {Battersby}
  C.~D., {Clark} P.~C., {Hatchfield} H.~P., {Smith} R.~J., 2020, \mnras, 499,
  4455

\bibitem[{{Tsujimoto} \& {Baba}(2019)}]{TsujimotoBaba2019}
{Tsujimoto} T., {Baba} J., 2019, \apj, 878, 125

\bibitem[{{Vasiliev}(2019)}]{Vasiliev2019}
{Vasiliev} E., 2019, \mnras, 482, 1525

\bibitem[{{Verley} {et~al}\mbox{.}(2007){Verley}, {Combes},
  {Verdes-Montenegro}, {Bergond}, \& {Leon}}]{Verley+2007}
{Verley} S., {Combes} F., {Verdes-Montenegro} L., {Bergond} G., {Leon} S.,
  2007, \aap, 474, 43

\bibitem[{{Vislosky} {et~al}\mbox{.}(2024){Vislosky}, {Minchev}, {Khoperskov},
  {Martig}, {Buck}, {Hilmi}, {Ratcliffe}, {Bland-Hawthorn}, {Quillen},
  {Steinmetz}, \& {de Jong}}]{Vislosky+2024}
{Vislosky} E. {et~al.}, 2024, \mnras, 528, 3576

\bibitem[{{Wegg} \& {Gerhard}(2013)}]{WeggGerhard2013}
{Wegg} C., {Gerhard} O., 2013, \mnras, 435, 1874

\bibitem[{{Wegg}, {Gerhard} \& {Portail}(2015){Wegg}, {Gerhard}, \&
  {Portail}}]{Wegg+2015}
{Wegg} C., {Gerhard} O., {Portail} M., 2015, \mnras, 450, 4050

\bibitem[{{Wylie}, {Clarke} \& {Gerhard}(2022){Wylie}, {Clarke}, \&
  {Gerhard}}]{Wylie+2022}
{Wylie} S.~M., {Clarke} J.~P., {Gerhard} O.~E., 2022, \aap, 659, A80

\bibitem[{{Zhang} {et~al}\mbox{.}(2025){Zhang}, {Belokurov}, {Evans},
  {Sanders}, {Lu}, {Cao}, {Myeong}, {Dillamore}, {Kane}, \& {Li}}]{Zhang+2025}
{Zhang} H. {et~al.}, 2025, \apjl, 983, L10

\end{thebibliography}

\end{document}